\documentclass[cits]{PoS}

\title{Synthetic Observations of the H\,I Line in SPH-Simulated Spiral Galaxies}

\ShortTitle{Synthetic H\,I Observations of SPH Galaxies}

\author{\speaker{Kevin A.~Douglas}\\%
       University of Exeter\\
       E-mail: \email{douglas@astro.ex.ac.uk}}

\author{David Acreman, Clare Dobbs, Chris Brunt\\
        University of Exeter}  

\abstract{Using the radiative transfer code Torus, we produce spectral-line cubes of the
predicted H\,I profile from global SPH simulations of spiral galaxies.  Torus grids the SPH
galaxy using Adaptive Mesh Refinement, then applies a ray-tracing method to infer the H\,I
profile along the line(s) of sight.  The gridded galaxy can be observed from any direction,
which enables us to model the observed H\,I profile for galaxies of any orientation.  We can
also place the observer inside the galaxy, to simulate H\,I observations taken from the Earth's
position in the Milky Way.}

\FullConference{Panoramic Radio Astronomy: Wide-field 1-2 GHz research on galaxy evolution\\
		 June 2-5 2009\\
		 Groningen, the Netherlands}

\begin{document}

\section{Introduction}

A significant difference between observations and numerical simulations of the interstellar medium is
that while observers detect emission through spectral line or continuum processes, often using
frequency/wavelength information to convert to radial velocity scales, numerical models directly
trace the total density of interstellar matter, with no reliance on kinematics to assign distances
to features seen in the simulations.  We seek to bridge that gap, using the Torus code [1]
which can apply radiative transfer to a gridded simulation (in our case an entire galaxy) and
predict the emission from selected tracer species.  

\section{Torus and the 21-cm line}

As one of the best tracers of interstellar and galactic structure,
the 21-cm neutral hydrogen (H\,I) spectral line is the first species we have implemented into
our analysis.  The simplicity of the detailed balance governing this two-level hyperfine transition
enables the easy calculation of important parameters such as the emissivity and opacity of the line,
and its profile function.  The relevant physical variables needed to perform these calculations are 
the density (of H\,I), temperature and velocity of the gas.  

In our work, Torus grids an SPH-simualted spiral galaxy onto an Adaptive Mesh Refinement (AMR) 
grid, essentially
transforming the particle view of the galaxy into a gridded view.  The galaxy model we have used
comes from a simulation investigating molecular cloud formation in spiral galaxies by [2], based on the 
Milky Way and containing a full thermodynamical model 
of the ISM, and a consistent treatment of the formation of molecular hydrogen.
Once the galaxy is gridded, the Torus code can manipulate the grid spatially, and perform radiative
transfer and ray tracing calculations to infer a synthetic line profile from an arbitrary observer
position.  Tracing multiple lines of sight can generate an image, or a three-dimensional cube of the
selected tracer species in the case of spectral lines, which is directly comparable to real observations.

To simulate extragalactic H\,I observations of (nearby) spiral galaxies, we can orient the SPH galaxy
to arbitrary position and inclination angles, and assign to it some systemic recessional or approach
velocity, so that the simulated datacube may be compared directly to typical radioastronomical 
observational data.  

\section{M33 as a test case}

As a demonstration of our ability to produce synthetic spectral datacubes that successfully imitate the
observed H\,I profile, we chose the nearby spiral galaxy M33, primarily because of the availability
of high-quality observational data [3] with which our results could be compared.  Quantitative 
differences are to be expected, given the circular rotation of M33 is approximately 100 km/s [4] while
the SPH galaxy, modelled to be more akin to the Milky Way, has $v_{rot} \approx 220$ km/s.  Moreover,
the interaction of M33 with its surroundings (especially M31) gives rise to extended H\,I emission
well beyond its main disk, and we would therefore not expect even a qualitative match to the SPH galaxy
in this component.

\begin{figure}
\begin{center}
\includegraphics[width=.75\textwidth]{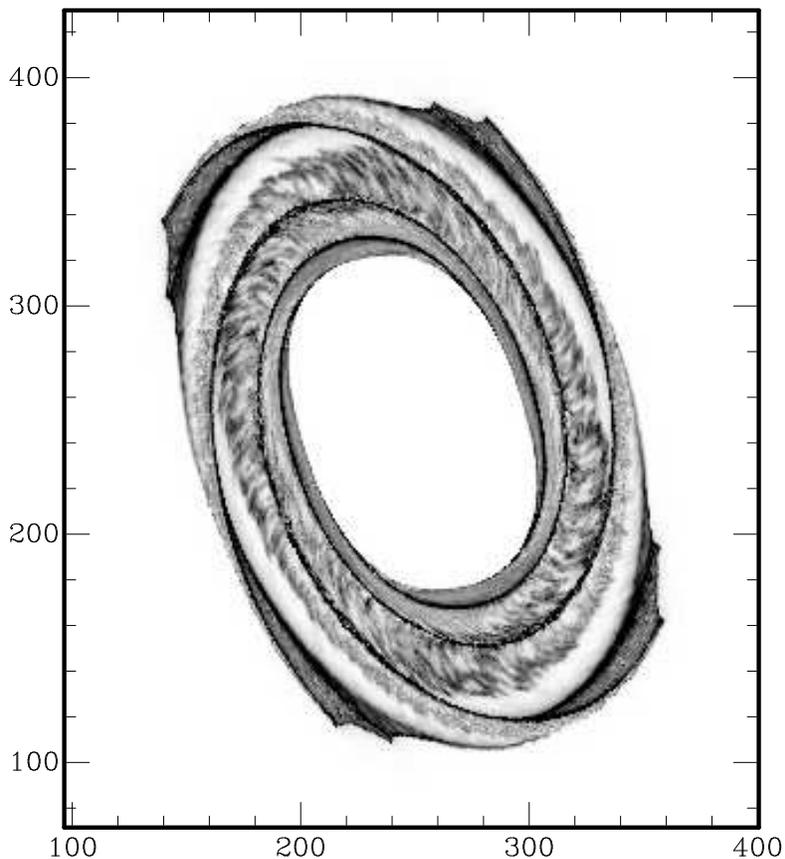}
\end{center}
\caption{Column Density image of the SPH galaxy, oriented to match M33's position and inclination
angles.   Note that the SPH calculation does not trace gas within 5 kpc of the galaxy's center.
The axes reflect the datacube pixel values.}
\label{fig1}
\end{figure}

In Figure \ref{fig1} we show a total column density image of the
SPH galaxy, oriented to match the disk of M33.  
Numerically, we can resolve the structure in the SPH
galaxy better than the angular resolution of the M33 data, so we convolve our synthetic datacube with
a Gaussian beam to simulate the observations being done by a single-dish telescope.  

\begin{figure}
\includegraphics[width=.5\textwidth]{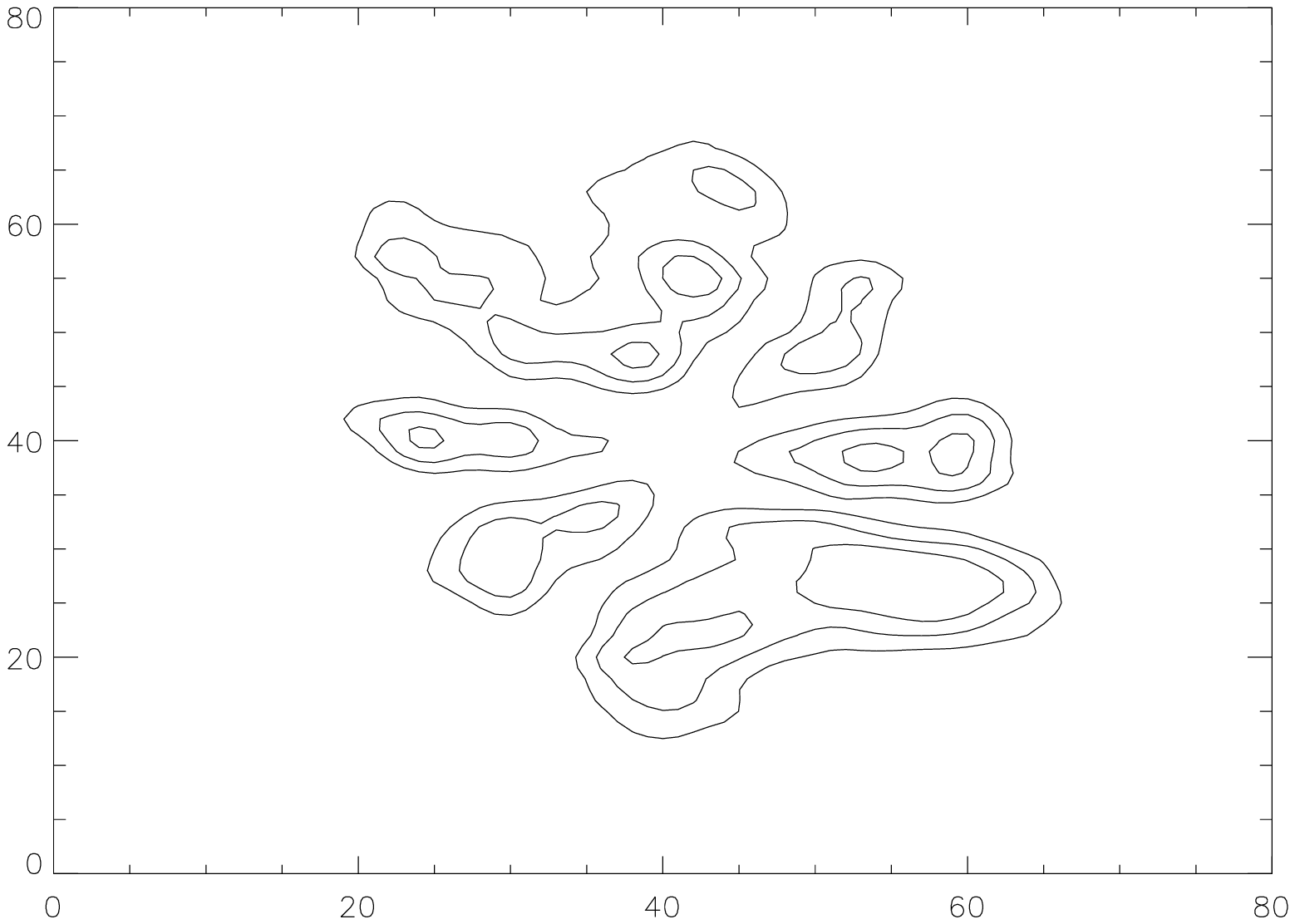}
\includegraphics[width=.5\textwidth]{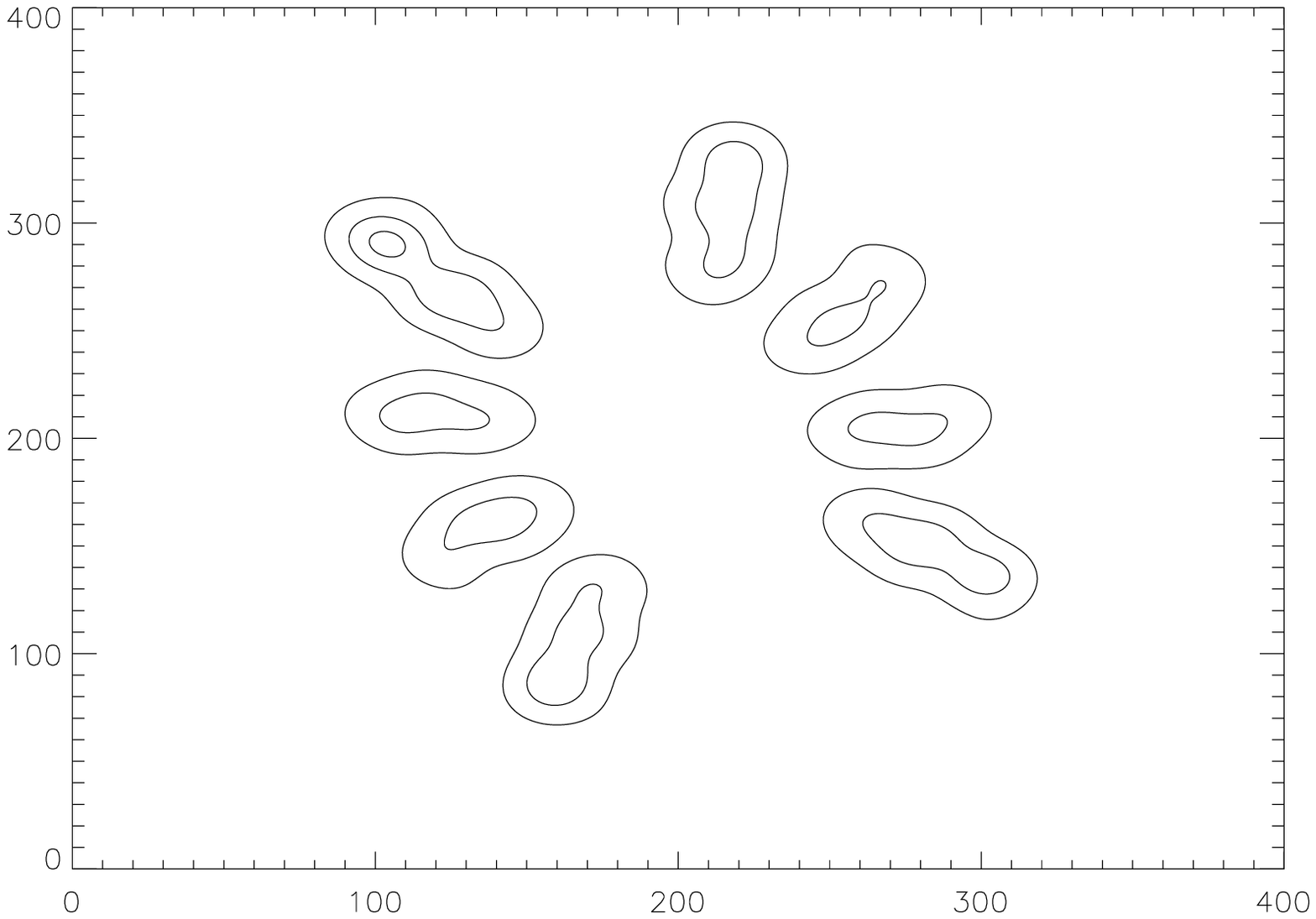}
\caption{Isovelocity contour plots of M33 (left) and the SPH galaxy (right).  The four contours are from
different channel maps of the H\,I data cube, with the top contour showing the most blueshifted H\,I and
progressing to the most redshifted H\,I in the bottom contours.  Axes reflect the datacube pixel scales.}
\label{fig2}
\end{figure}

The qualitative agreement we seek is the basic, expected observation of the galaxy's main disk being
traced by a series of isovelocity curves as we step through the cube in radial velocity.  We show
four such curves as contour plots for the M33 cube in the left panel of figure \ref{fig2}.
Our results in the right panel of figure \ref{fig2} show a good qualitative
match to the general rotational structure of M33.  

The structure contained in the simulated cube at the native resolution of the simulations
hints at the effects of spiral shocks and other noncircular motions
contained within the rotating galaxy disk.  Very high resolution observations with interferometric 
radio telescopes, and eventually the SKA, will make such investigations of fine structure in the spiral
arms of galaxies routine.

\section{Conclusion and Future Directions}

We have described a model to transform output from numerical simulations into synthetic H\,I datacubes, 
which could be measured by an `observer' inside, or outside a galaxy.  For external observers, the case
most relevant to this conference, we can 
produce maps of higher resolution than available with current instruments.  Within a galaxy, this process 
has the advantage that distances to features are known precisely, rather than being subject to large 
kinematic uncertainties.

As with all situations, this method is only as good as the model.  We have begun with an SPH simulation
of a spiral galaxy which emulates many of the aspects of the Milky Way.  However, important physical
processes such as stellar feedback have not yet been implemented, and so certain quantitative matches
have not yet been achieved.  Nevertheless this line of work may prove valuable to the study of spiral
structure in galaxies, especially as instruments leading to the Square Kilometre Array continue to push
the limits of angular resolution.

\acknowledgments
The research leading to these results has received funding from the Seventh Framework Programme under 
grant agreement n$^{\rm o}$ PIIF-GA-2008-221289.  KAD wishes to thank the PRA conference organisers for a
thoroughly enjoyable and forward-looking experience in Groningen.

\end{document}